\documentclass[aps,prb,twocolumn,showpacs]{revtex4}

\usepackage{graphicx}

\begin{document}

\title{Ground-state structure of KNbO$_3$/KTaO$_3$ superlattices: \\
Array of nearly independent ferroelectrically ordered planes}

\author{Alexander I. Lebedev}
\email[]{swan@scon155.phys.msu.ru}
\affiliation{Physics Department, Moscow State University, \\
Leninskie gory, 119991 Moscow, Russia}

\date{\today}

\begin{abstract}
Phonon spectra of the paraelectric phase, structure and distribution of
polarization in the polar ground state, and energies of ferroelectrically
and antiferroelectrically ordered phases were calculated for stress-free
(KNbO$_3$)$_1$(KTaO$_3$)$_n$ superlattices ($n = {}$1--7) from first
principles within the density-functional theory. Quasi-two-dimensional
ferroelectric state with the polarization localized in the KNbO$_3$ layer
was revealed to be the ground
state for the superlattices with $n \ge 2$. The interaction between these
polarized layers decreases exponentially with increasing the interlayer
distance, and the ground-state structure transforms to an array of nearly
independent ferroelectrically ordered planes at large $n$.
\end{abstract}


\maketitle

\section{Introduction}

Ferroelectric superlattices (SLs) have drawn much attention in recent years.
It appeared that many physical properties of these artificial structures,
such as the Curie temperature, spontaneous polarization, dielectric constant,
nonlinear dielectric and optical susceptibilities, exceed considerably the
properties of bulk crystals and thin films of the corresponding solid
solutions.

Ferroelectric phenomena in KTa$_{1-x}$Nb$_x$O$_3$ solid solutions and
KNbO$_3$/KTaO$_3$ superlattices have been studied for many years. The
experiment shows that in solid solutions a polar phase appears at
$x > x_c = 0.008$,~\cite{PhysRevLett.39.1158} but its properties in the
vicinity of critical concentration $x_c$ are modified by manifestation
of quantum effects and by local disorder (the dipolar glass). To explain the
peculiar properties of the solid solutions, a number of models has been
proposed~\cite{PhysRevB.23.221,PhysRevB.33.2084} including a hypothesis
about off-centering of Nb atoms in
KTaO$_3$.~\cite{PhysRevLett.53.298,PhysRevB.44.6700,JPhysCondensMatter.13.8817,
Ferroelectrics.164.101,JPhysCondensMatter.10.6271} However, as was shown
in Ref.~\onlinecite{PhysSolidState.44.1135}, the single-well shape of the local
potential for Nb atom in the solid solution excludes its off-centering.

Ferroelectric properties of KNbO$_3$/KTaO$_3$ superlattices have been studied
both experimentally~\cite{ApplPhysLett.68.1488,ApplPhysLett.72.2535,
PhysRevLett.80.4317,JElectroceram.4.279,PhysRevLett.88.097601,
JVacSciTechnolA.22.2010} and
theoretically.~\cite{PhysRevB.64.060101,JApplPhys.90.4509,JApplPhys.91.3165,
ApplPhysLett.86.232903,PhysSolidState.52.1448,cond-mat.0401039}
The ferroelectric properties of the SLs grown on KTaO$_3$ substrates were
calculated assuming that the polar ground-state structure is
tetragonal.~\cite{ApplPhysLett.86.232903}
The signs of antiferroelectricity in these SLs observed
experimentally~\cite{PhysRevLett.88.097601} were confirmed by observation
of unstable phonon at $X$ point of the Brillouin zone in first-principles
calculation.~\cite{ApplPhysLett.86.232903}  At the same time, our recent
calculations of the ground state for both stress-free and substrate-supported
superlattices with $n = 1$ and 3 (Ref.~\onlinecite{PhysSolidState.52.1448})
revealed that the ground-state structure of KNbO$_3$/KTaO$_3$ superlattices is
monoclinic or orthorhombic, and is not tetragonal. Along with the ferroelectric
instability, three unstable phonon modes that are not typical neither for
KNbO$_3$ nor for KTaO$_3$ were observed at $X$, $Z$, and $R$ points of the
Brillouin zone in the paraelectric phase of these SLs. In contrast to other
instabilities found in ten SLs studied in
Ref.~\onlinecite{PhysSolidState.52.1448}, the origin of these instabilities
was not clear. The calculations~\cite{cond-mat.0401039}
of the phonon frequencies at the $\Gamma$ point for the paraelectric phase of
(KNbO$_3$)$_1$(KTaO$_3$)$_7$ superlattice with the lattice parameter equal
to that of KTaO$_3$ revealed a weak ferroelectric
instability of this phase with the polarization parallel to the layer plane.
However, the distribution of polarization in the low-symmetry polar phase
was not analyzed.

The purpose of this paper was to study the development of the ferroelectric
properties in stress-free (KNbO$_3$)$_1$(KTaO$_3$)$_n$ superlattices when
increasing their period~$L = n + 1$. We show that quasi-two-dimensional (2D)
ferroelectric state with polarization localized in the KNbO$_3$ layer is
the ground-state structure for the superlattices with $n \ge 2$ and it
transforms to an array of nearly independent ferroelectrically ordered planes
at large $n$.

\section{Calculation details}

The (KNbO$_3$)$_1$(KTaO$_3$)$_n$ superlattices considered in this work are
periodic structures grown in [001] direction and consisted of the KNbO$_3$
layer with a thickness of one unit cell and the KTaO$_3$ layer with a thickness
of $n$~unit cells ($1 \le n \le 7$). The modeling of these structures was
performed using supercells with a size of 1$\times$1$\times$$(n + 1)$ unit
cells.

The calculations were performed within the first-principles density-functional
theory (DFT) with pseudopotentials and a plane-wave basis set as implemented
in \texttt{ABINIT} software.~\cite{abinit}  The local density
approximation (LDA)~\cite{PhysRevB.23.5048} for the exchange-correlation
functional was used. Optimized separable nonlocal
pseudopotentials~\cite{PhysRevB.41.1227} were constructed using the
\texttt{OPIUM} software;~\cite{opium} to improve the transferability of
pseudopotentials, the local potential correction~\cite{PhysRevB.59.12471} was
added.  Parameters used for construction of pseudopotentials and other details
of calculations are given in
Ref.~\onlinecite{PhysSolidState.52.1448}. The plane-wave cut-off energy was
40~Ha (1088~eV). The integration over the Brillouin zone was performed on a
8$\times$8$\times$4 Monkhorst-Pack mesh for SLs with $n = 1$ and 2, and on a
8$\times$8$\times$2 one for SLs with $n = {}$3--7. The relaxation of atomic
positions and of the unit cell parameters was stopped when the Hellmann-Feynman
forces were below $5 \cdot 10^{-6}$~Ha/Bohr (0.25~meV/{\AA}). The phonon
spectra were calculated using the density-functional perturbation theory. The
spontaneous polarization was calculated using the Berry phase
method.~\cite{PhysRevB.47.1651}

The structures of KNbO$_3$ and KTaO$_3$ calculated using this approach agree
well with experiment. The calculated lattice parameters for cubic KNbO$_3$ and
KTaO$_3$ are 3.983 and 3.937~{\AA}, they differ from the experimental values
(4.016 and 3.980~{\AA}) by 0.81 and 1.07\%, respectively (small underestimation
of the lattice parameters is typical for LDA approximation). The calculated
$c/a$ ratio of 1.0197 for tetragonal KNbO$_3$ is close to the experimental
value (1.0165). The ground-state structure is rhombohedral ($R3m$) for KNbO$_3$
and cubic ($Pm3m$) for KTaO$_3$, in agreement with experiment. The calculated
spontaneous polarization of 0.372~C/m$^2$ for tetragonal KNbO$_3$ is close to
experimental values of 0.37--0.39~C/m$^2$ (Ref.~\onlinecite{PhysRevB.30.1148}).

\section{Results}

\begin{figure}
\centering
\includegraphics{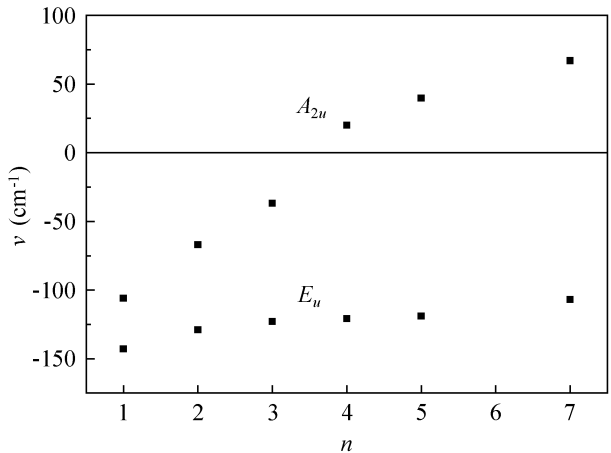}
\caption{Frequencies of lowest
$A_{2u}$ and $E_u$ polar optical modes in the paraelectric $P4/mmm$
phase as a function of the thickness of the KTaO$_3$ layer in
(KNbO$_3$)$_1$(KTaO$_3$)$_n$ superlattices. Imaginary frequencies of
unstable modes are plotted as negative values.}
\label{fig1}
\end{figure}

Calculated frequencies of two lowest $A_{2u}$ and $E_u$ ferroelectric modes
at the $\Gamma$ point of the Brillouin zone in the paraelectric $P4/mmm$ phase
of stress-free (KNbO$_3$)$_1$(KTaO$_3$)$_n$ superlattices with $n = {}$1--7
are presented in Fig.~\ref{fig1}. In SL with $n = 1$ both the modes, which
result from unstable TO phonon of the perovskite structure, are unstable.
With increasing $n$, the $A_{2u}$ mode becomes stable, and there remains
one unstable $E_u$ phonon mode, whose frequency is weakly dependent on the
KTaO$_3$ layer thickness.

When searching for an equilibrium structure, all possible structure distortions
were tested. For SLs with $n = 1$, 2, and 3, in which two unstable modes
appeared in the paraelectric phase, monoclinic phases with $Cm$ and $Pm$ space
groups (polarized along [$xxz$] and [$x0z$] directions) as well as tetragonal
$P4mm$ and two orthorhombic $Amm2$ and $Pmm2$ phases were considered. For SLs
with $n > 3$ only the $E_u$ mode is unstable, and so only orthorhombic $Amm2$
and $Pmm2$ phases polarized along [$xx0$] and [$x00$] directions were possible.
The calculations revealed that for SLs with $n = 1$ the ground-state structure
is $Cm$~\cite{PhysSolidState.52.1448} whereas for other SLs---in spite of the
existence of unstable $A_{2u}$ mode in the paraelectric phase of SLs with
$n = 2$ and 3---the ground-state structure is $Amm2$.

\begin{table}
\caption{\label{table2}Frequencies of unstable phonon modes at high-symmetry
points of the Brillouin zone in the $P4/mmm$ phase of (KNbO$_3$)$_1$(KTaO$_3$)$_n$
superlattices with $n = 1$, 2, and 3.}
\begin{ruledtabular}
\begin{tabular}{ccccc}
$n$ & \multicolumn{4}{c}{$\nu$ (cm$^{-1}$)} \\
\cline{2-5}
    & $\Gamma$~(0,0,0) & $Z$~(0,0,1/2) & $X$~(1/2,0,0) & $R$~(1/2,0,1/2) \\
\hline
1   & 143$i$ & 126$i$ & 71$i$ & 61$i$ \\
2   & 118$i$ & 114$i$ & 50$i$ & 50$i$ \\
3   & 114$i$ & 113$i$ & 46$i$ & 45$i$ \\
\end{tabular}
\end{ruledtabular}
\end{table}

Calculation of phonon frequencies at high-symmetry points of the
Brillouin zone for paraelectric $P4/mmm$ phase of SLs with $n \le 3$
reveals, in addition to the ferroelectric instability, unstable phonons
at $Z$, $X$, and $R$ points (Table~\ref{table2}). However, among all
possible distorted structures the polar $Cm$ and $Amm2$ phases had the
lowest total energy, and all phonons at all points of the Brillouin zone
were stable in these phases.

\begin{table*}
\caption{\label{table1}The total energy difference between paraelectric and
ground-state structures, the lattice parameters, and the Berry-phase
polarization for the ground-state structure of (KNbO$_3$)$_1$(KTaO$_3$)$_n$
superlattices with different $n$.}
\begin{ruledtabular}
\begin{tabular}{ccccccc}
$n$              & 1       & 2       & 3       & 4       & 5       & 7 \\
\hline
$\Delta E$ (meV) & 11.738  &  7.532  &  5.898  &  5.365  &  5.090  & 4.625 \\
$a = b$ ({\AA})  &  3.9656 &  3.9567 &  3.9508 &  3.9478 &  3.9459 & 3.9437 \\
$c$ ({\AA})      &  7.9235 & 11.8500 & 15.7908 & 19.7298 & 23.6680 & 31.5436 \\
$\gamma$ (deg.)  & 89.9581 & 89.9746 & 89.9866 & 89.9913 & 89.9936 & 89.9958 \\
$\alpha=\beta$ (deg.) & 89.9701 & 90.0 & 90.0  & 90.0    & 90.0    & 90.0 \\
$|P_s|$ (C/m$^2$) & 0.2515 &  0.1192 &  0.0822 &  0.0612 &  0.0486 &  0.0338 \\
\end{tabular}
\end{ruledtabular}
\end{table*}

As follows from Table~\ref{table1}, the energy gain resulting from
the ferroelectric ordering does not vanish with increasing $n$. This fact
and the existence of unstable $E_u$ phonon in the paraelectric
phase give evidence for stability of the ferroelectric state in KNbO$_3$
layers of minimal thickness (one unit cell) located between layers of
KTaO$_3$ of arbitrary thickness.

\begin{figure}
\medskip
\centering
\includegraphics{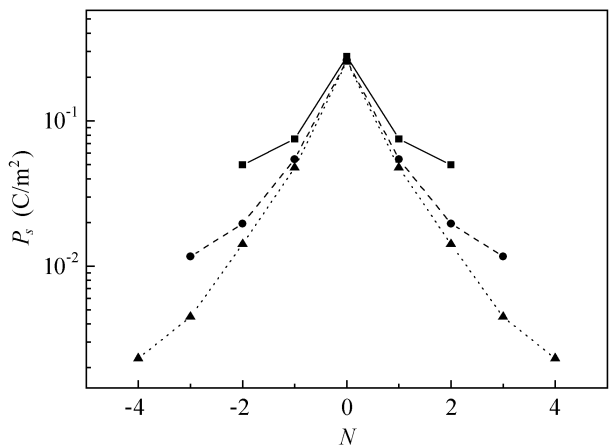}
\caption{The polarization profiles for (KNbO$_3$)$_1$(KTaO$_3$)$_n$
superlattices with $n = {}$3, 5, and 7. The number $N = 0$ corresponds to
the KNbO$_3$ layer.}
\label{fig2}
\end{figure}

The polarization profiles for the superlattices were calculated using the
approximate formula
$P_{s\alpha} = (e / \Omega) \sum_i w_i Z^*_{i,\alpha\beta} u_{i\beta}$,
where $P_{s\alpha}$ is the $\alpha$-component of polarization ($\alpha = x,
y, z$), $\Omega$ is the unit cell volume of the layer under consideration,
$u_{i\beta}$ is the $\beta$-component of displacement of the $i$th atom
from its position in the paraelectric phase, $Z^*_{i,\alpha\beta}$ is the
Born effective charge tensor of this atom,%
    \footnote{Born effective charge tensors obtained for $P4/mmm$ phase of
    the superlattice with $n = 1$ were used in calculations; the effective
    charges for SLs with $n = {}$2--7 were very close to those used.}
and $w_i$ is the weight equal to 1 for Nb(Ta) and O atoms lying in the
Nb(Ta)--O layer under consideration, 1/2 for K and O atoms lying in two
nearest K--O planes, and 0 for other atoms. The dependence of $|P_s|$ on the
layer number in SLs with $n = 3$, 5, and 7 are given in Fig.~\ref{fig2}.
It is seen that in the region between KNbO$_3$ layers the polarization
decays exponentially, with a characteristic length of $\sim$3~{\AA}. This
result indicates a strong localization of polarization in the potassium niobate
layer and proves the \emph{quiasi-two-dimensional character of ferroelectricity}
in considered superlattices. The polarization in the KNbO$_3$ layer decreases
monotonically from 0.279 to 0.257~C/m$^2$ with increasing $n$ from 3 to 7
(for comparison, for bulk orthorhombic KNbO$_3$ our calculations give
$|P_s| = 0.418$~C/m$^2$). The total polarization of the supercells calculated
using the above formula differs from the result of correct Berry-phase
calculations (Table~\ref{table1}) by only 3\%.

\begin{table}
\caption{\label{table3}The energy difference between ferroelectrically and
antiferroelectrically ordered structures in [(KNbO$_3$)$_1$(KTaO$_3$)$_n$]$_2$
superlattices (per Nb atom).}
\begin{ruledtabular}
\begin{tabular}{cccc}
$n$                  & 1     & 2     & 3 \\
\hline
$2W_{\rm int}$ (meV) & 4.282 & 1.089 & 0.278 \\
\end{tabular}
\end{ruledtabular}
\end{table}

In order to determine the interaction energy $W_{\rm int}$ between two
polarized KNbO$_3$ layers separated by KTaO$_3$ layer, the total energies
of [(KNbO$_3$)$_1$(KTaO$_3$)$_n$]$_2$ superlattices of doubled period with
ferroelectric and antiferroelectric ordering of polarization in neighboring
KNbO$_3$ layers were calculated. The energy difference between the two ordered
structures (per Nb atom) is given in Table~\ref{table3}. As follows from
these data, the energy difference decreases exponentially with increasing
thickness of KTaO$_3$ layer with a charactricstic decay length of 2.9~{\AA}.

The polarization in two-dimensional layer is stable against its spontaneous
reversal if $2W_{\rm int} < U$, where $W_{\rm int}$ is the energy of
interlayer interaction and $U$ is the height of the potential barrier
separating different orientational states of polarization in the layer. The
multiplier two in the above formula relates to the worst case in which the
considered layer is polarized antiparallel to the polarizations in neighboring
layers. In our structures, the easiest way to reorient the polarization is
to rotate it in the layer plane, and so the $U$ value is equal to the energy
difference between [110]- and [100]-polarized structures. According to our
calculations, $U = 1.84$~meV for SL with $n = 2$ and 1.69~meV for SL with
$n = 3$. Comparison of $U$ and $2W_{\rm int}$ values (Table~\ref{table3})
shows that the stability criterion is satisfied at $n \ge 2$. From the data
presented in Tables~\ref{table1} and \ref{table3} one can see that in
superlattices with $n \ge 3$ the interaction energy between neighboring
polarized layers is less that 10\% of the ordering energy. This gives the
reason to consider the ground state in these superlattices as \emph{an array
of nearly independent ferroelectrically ordered planes}, with
quasi-two-dimensional ferroelectricity realized in each layer.

The results obtained for (KNbO$_3$)$_1$(KTaO$_3$)$_n$ superlattices grown on
KTaO$_3$ substrates are very different from those obtained for stress-free
superlattices. As follows from Table~\ref{table4}, for KTaO$_3$-supported
superlattices the ground-state structure is monoclinic for $n \le 4$ and
orthorhombic for $n = {}$5--7 (data for $n = 6$ and 7 are not shown). This
result can be understood as follows. For KNbO$_3$ grown on KTaO$_3$ substrate,
the ground-state structure is $Cm$. The uniform $P_x$-component of polarization
does not create macroscopic electric field in SLs, so that the interaction
of polarizations in neighboring layers is weak and $P_x$ varies slowly with
increasing $n$. On the contrary, the energy needed to maintain the same
electric displacement fields normal to the layers in two materials increases
with increasing the thickness of nonpolar layer $n$, thus weakening the
instability of $z$-polarized unstable phonon. This results in fast decrease
of $P_z$ with increasing $n$.

Another interesting result that follows from Table~\ref{table4} is an
increase of the in-plane polarization in the KNbO$_3$ layer with increasing
$n$. At first sight, this result seems strange because the in-plane lattice
parameter in SLs remains unchanged. However, if one takes into account the
influence of $P_z$ on the effective charge of the $E_u$ soft mode, the result
can be explained. According to our calculations, in strained KNbO$_3$ grown
on KTaO$_3$ substrate the effective charge decreased by 7.7\% when the
structure transformed from paraelectric $P4/mmm$ to polar $P4mm$ phase.
Thus, when increasing $n$, the decrease in $P_z$ results in an increase of
effective charge of $E_u$ mode. This strengthens the dipole-dipole
interaction and increases the instability of the soft mode.

\begin{table}
\caption{\label{table4}The ground-state structure, lattice parameters, the
Berry-phase polarization ($P_x$, $P_y$, $P_z$), and the in-plane polarization
in the KNbO$_3$ layer ($P_x^*$, $P_y^*$) for (KNbO$_3$)$_1$(KTaO$_3$)$_n$
superlattices grown on KTaO$_3$ substrate.}
\begin{ruledtabular}
\begin{tabular}{cccccc}
$n$                   & 1      &  2      &  3      &  4      &  5 \\
\hline
Structure             & $Cm$   &  $Cm$   &  $Cm$   &  $Cm$   &  $Amm2$ \\
$a = b$ ({\AA})       & 3.9374 &  3.9374 &  3.9374 &  3.9374 &  3.9374 \\
$c$ ({\AA})           & 7.9903 & 11.9126 & 15.8358 & 19.7591 & 23.6861 \\
$P_x = P_y$ (C/m$^2$) & 0.0730 &  0.0605 &  0.0507 &  0.0442 &  0.0385 \\
$P_z$ (C/m$^2$)       & 0.2546 &  0.1769 &  0.1216 &  0.0710 &  0.0000 \\
$P_x^* = P_y^*$ (C/m$^2$) & 0.1322 & 0.1495 & 0.1585 & 0.1646 & 0.1675 \\
\end{tabular}
\end{ruledtabular}
\end{table}

\section{Discussion}

The critical size of ferroelectric particles and critical thickness of
ferroelectric thin films, below which the ferroelectricity vanishes, are
of principle importance. First studies of ferroelectric nanoparticles
indicated that the ferroelectricity vanishes when the particle diameter is
about 100~{\AA}. However, the experiments on ultrathin ferroelectric
films~\cite{ApplPhysLett.75.856} and theoretical calculations from first
principles~\cite{ApplPhysLett.76.2767,PhysRevB.63.205426,Nature.422.506,
PhysRevB.72.020101,PhysRevB.74.060101} have shown that a non-zero
polarization normal to the film surface is retained in films with a thickness
of 3--6 unit cells if the depolarizing field is compensated. In the
PVDF-TrFE copolymer films the switchable polarization was observed in
films with a thickness of 10~{\AA} (two monolayers).~\cite{Nature.391.874}
The ferroelectric state with the polarization \emph{parallel} to the layers
was predicted for BaTiO$_3$ and PbTiO$_3$ films with a thickness of
3~unit cells~\cite{PhysRevB.56.1625, FaradayDiscuss.114.395} and PZT films
with a thickness of one unit cell.~\cite{PhysRevB.70.220102}
Unstable ferroelectric mode with the polarization in the film layer was
predicted for films of Pb$B^\prime_{1/2}B^{\prime\prime}_{1/2}$O$_3$ solid
solutions with a thickness of one unit cell.~\cite{PhysSolidState.51.1894}
In all these cases two-dimensional ferroelectricity was a result of
decreasing the physical size of the samples in one direction. In contrast,
in this work, quasi-two-dimensional ferroelectricity appears as a result of
specific interactions in superlattices and can be obtained in structures of
arbitrary thickness.

The polarization profiles obtained in this work agree with results of
atomistic modeling of KNbO$_3$/KTaO$_3$
superlattices.~\cite{PhysRevB.64.060101,JApplPhys.90.4509}
According to the data of these papers, the polarization parallel to the
layers decreased by 3--4 times when shifting by one unit cell from the
heterojunction (in contrast to the behavior of the polarization normal to
the layers). Our conclusion about the instability of $A_{2u}$ phonon in
the paraelectric phase of SLs
with $n \le 3$ agrees with results of Ref.~\onlinecite{ApplPhysLett.86.232903}
only partially: in the latter paper stable polar phase with $P4mm$ symmetry
was obtained only in SLs with $n = 1$ and 2. The instability of the $E_u$
mode in SL with $n = 7$ agrees with that observed in
Ref.~\onlinecite{cond-mat.0401039}, but in this work it is much stronger.

However, in spite of qualitative agreement of results of this and previous
works, the physical conclusions that we make are very different. First, in
all stress-free KNbO$_3$/KTaO$_3$ superlattices the $P4mm$ phase is not the
ground-state structure, as well as in KTaO$_3$-supported superlattices with
$n \ge 5$. The reason for this is a tendency of polarization in SLs to
incline towards the layer plane. This tendency was revealed in a number of
ferroelectric superlattices~\cite{PhysSolidState.52.1448} and was shown to
be a way to reduce the electrostatic and mechanical energies in them.

Second, the existence of stable polarization in KNbO$_3$ layers with a
thickness of one unit cell and very inhomogeneous distribution of
polarization, which is accompanied by exponential decay of the interaction
energy between neighboring layers with increasing the interlayer distance,
result in a new, previously unknown feature of the ground-state structure of
(KNbO$_3$)$_1$(KTaO$_3$)$_n$ superlattice---the formation of an array of
nearly independent ferroelectrically ordered planes in the bulk of the
superlattice.

\begin{figure}
\includegraphics{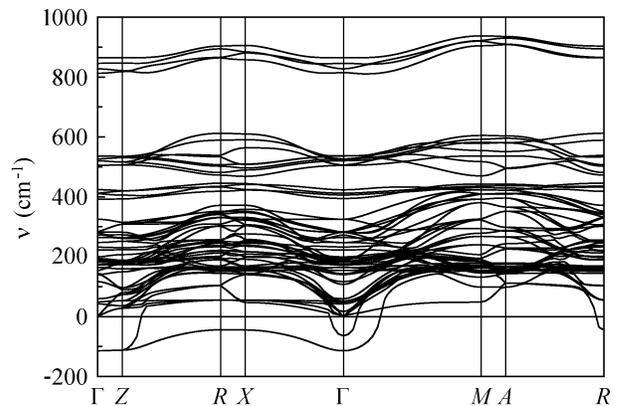}
\caption{\label{fig3}Phonon dispersion curves for paraelectric $P4/mmm$ phase
of (KNbO$_3$)$_1$(KTaO$_3$)$_3$ superlattice.}
\end{figure}

Phonon dispersion curves calculated using the interpolation technique from
phonon frequencies at $\Gamma$, $X$, $M$, $Z$, $R$, and $A$ points of the
Brillouin zone for $P4/mmm$ phase of (KNbO$_3$)$_1$(KTaO$_3$)$_n$ superlattices
enables to clarify the origin of unstable phonon modes at the boundary of the
Brillouin zone in these SLs.~\cite{ApplPhysLett.86.232903,PhysSolidState.52.1448}
The dispersion curves for these modes are shown in Fig.~\ref{fig3}. Analysis
of the eigenvectors shows that the doubly degenerate unstable mode along
$\Gamma$--$Z$ direction is polarized along [100] and [010] directions while
nondegenerate unstable modes along $Z$--$R$, $R$--$X$, and $X$--$\Gamma$
directions are polarized along [010] direction. The character of atomic
displacements for these modes is fully consistent with the ferroelectric
instability of ...--O--Nb--O--... chains first observed in
KNbO$_3$,~\cite{PhysRevLett.74.4067}  but in SLs the chains are formed only
in [100] and [010] directions. A~nearly zero
dispersion of the unstable mode along $\Gamma$--$Z$ direction means that the
interlayer interaction is weak, and the instability against the formation of
ferroelectrically ordered planes does not depend on whether neighboring planes
are ordered ferroelectrically or antiferroelectrically. This supports the idea
about independence of ferroelectrically ordered planes in SLs with thick
KTaO$_3$ layers.

The obtained results demonstrate that in complex systems in some cases the
polarization calculated using the Berry phase method cannot be interpreted
as a bulk property. The variations in polarization in the unit
cell can be so large that the ferroelectric system can acquire the properties
of two-dimensional system. To avoid interaction between polarized layers, the
polarization should be parallel to the layers (a non-zero normal component
results in appearance of electric fields in the structure, and 2D behavior
transforms to 3D one). This means that to create arrays of nearly independent
ferroelectrically ordered planes in [001]-oriented superlattices, we need
SLs with $Pmm2$ or $Amm2$ ground-state structure.

Complex ferroelectric/antiferroelectric behavior has been observed recently
in Ruddlesden--Popper (RP)
superlattices,~\cite{PhysRevB.78.064107,PhysRevB.82.045426} in which
appearance of Goldstone-like rotational mode of the in-plane polarization
is possible.~\cite{PhysRevLett.104.097601}  In stress-free BaTiO$_3$/SrO
superlattices, the in-plane polarization appeared in SLs with the BaTiO$_3$
slab thickness of $n \ge 4$ unit cells; when these SLs were epitaxially grown
on SrTiO$_3$ substrates, the polar structure transformed to antiferroelectric
(domain-like) structure in which out-of-plane polarization alternated along
[100] or [110] directions in each of BaTiO$_3$ slabs. The antiferroelectric
structure was studied in more detail in BaTiO$_3$/BaO superlattices grown
on SrTiO$_3$ substrate.~\cite{PhysRevB.82.045426}  The instability at $X$
point of the Brillouin zone appeared in structures at a thickness of BaTiO$_3$
slabs of 2--3 unit cells while the zone-center ferroelectric instability
appeared only at $n = {}$4--5 unit cells. It was supposed that a huge
depolarizing field appears in SLs because of the difficulty to polarize BaO
layer, and this strongly harden the $A_{2u}$ ferroelectric mode at the $\Gamma$
point. At the same time, the $X$ phonon mode arising from the same
ferroelectric instability of ...--O--Ti--O--... chains does not produce
macroscopic depolarizing field, and so the influence of the BaO layer on
the phonon frequency is weaker. This explains why in RP superlattices the
antiferroelectric phase can appear before the ferroelectric one.

The superlattices studied in this work differ markedly from RP
ones~\cite{PhysRevB.78.064107,PhysRevB.82.045426} because in our case the
ferroelectric layers are separated by the layers of incipient ferroelectric,
which has much higher dielectric constant compared to SrO (BaO). In contrast
to the RP superlattices, our SLs do not exhibit antiferrodistortive instability.
In addition, the thickness of the KNbO$_3$ layer which is necessary to
induce ferroelectricity in our SLs is only one unit cell. To test the
influence of the thickness of KNbO$_3$ layer on the appearance of
aforementioned antiferroelectric structure, we performed additional
calculations for stress-free SLs in which KNbO$_3$ layers with a thickness of
3 and 5~unit cells were separated with a thick (3~unit cells) KTaO$_3$ layer.

The calculations showed that all three (KNbO$_3$)$_n$(KTaO$_3$)$_3$ SLs with
$n = 1$, 3, and 5 studied in this work are unstable against the appearance of
antiferroelectric structure with out-of-plane displacements of ions described
by $X$ phonon ($Pmma$ space group), but in all cases the energies of
corresponding equilibrium structures are higher than the energies of the
ferroelectric $Amm2$ phases. This means that antiferroelectric $Pmma$ structure
is not the ground state of stress-free KNbO$_3$/KTaO$_3$ superlattice.
Moreover, the calculation of phonon frequencies at the $\Gamma$ point for
$Pmma$ phase of (KNbO$_3$)$_1$(KTaO$_3$)$_1$ SL showed that this
antiferroelectric phase exhibits the \emph{ferroelectric} instability against
the appearance of in-plane polarization. Therefore one can conclude that like for
RP superlattices, the antiferroelectric phase of KNbO$_3$/KTaO$_3$ superlattices
with out-of-plane displacements of ions is indeed more stable compared to
the ferroelectric phase with out-of-plane polarization, but nevertheless the
ferroelectric $Amm2$ phase with non-zero in-plane polarization has the lowest
energy. In RP superlattices studied so far, the in-plane polarization was
suppressed by epitaxial strain from the substrate, and the antiferroelectric
phase was the most stable one.

We discuss now possible applications of the arrays of nearly independent
ferroelectrically ordered planes. At first sight, the fact that the
polarization in KNbO$_3$/KTaO$_3$ superlattices is oriented in the layer plane
can be regarded as not intriguing. So far, when discussing the ferroelectricity
in ultrathin films, main attention was paid to films in which the polarization
is normal to the film surface as these films are more suitable for different
applications. However, the possibility of formation of an array of nearly
independent ferroelectrically ordered planes opens new opportunities in design
of electronic devices. These structures can be used as a medium for
three-dimensional information recording. Accepting that the lateral size of a
ferroelectric domain that has a long-time stability against spontaneous reversal
of polarization is 250~{\AA},~\cite{ApplPhysLett.79.530} for the interlayer
distance of 16~{\AA} (the period of SL with $n = 3$) the estimated 3D
information storage density can be as high as $\sim$$10^{18}$~bit/cm$^3$. This
value exceeds the density achieved in modern optical 3D recording media by six
orders of magnitude.

One can add that the two-component order parameter, which admits the rotation
of polarization by 90$^\circ$ in each of independent KNbO$_3$ layers, offers
one more opportunity. In ferroelectric RAM devices with polarization normal
to the film surface the leakage currents and conductivity of films are main
effects preventing the realization of non-destructive read-out. The systems
admitting the rotation of polarization by 90$^\circ$ have a feature that
enables to overcome this disadvantage---a non-destructive read-out of
polarization in them can be realized using the anisotropy of the dielectric
constant in the layer plane. Our calculations predict that in
(KNbO$_3$)$_1$(KTaO$_3$)$_n$ superlattice with $n = 2$ the dielectric
constants parallel and perpendicular to the polarization are 142 and 188,
whereas in the SL with $n = 3$ they are 225 and 281.

\section{Conclusions}

In this work, the density-functional theory was used to calculate the
properties of (KNbO$_3$)$_1$(KTaO$_3$)$_n$ superlattices with $n = {}$1--7.
It was shown that the ferroelectric ordering is retained in KNbO$_3$ layers
of minimal thickness (one unit cell) whereas the interaction energy between
neighboring polarized layers decreases exponentially with increasing
parameter $n$. Array of nearly independent ferroelectrically ordered
planes with the polarization lying in the layer plane was shown to be
the ground-state structure for SLs with $n \ge 2$. Using of these
structures as a medium for three-dimensional (3D) information recording
enables to attain the information storage density of $10^{18}$~bit/cm$^3$.

The calculations presented in this work were performed on the laboratory
computer cluster (16~cores) and SKIF-MGU ``Chebyshev'' supercomputer at
Moscow State University.

\begin{acknowledgments}
This work was supported by the Russian Foundation for Basic Research Grant
No.~08-02-01436.
\end{acknowledgments}

%\bibliography{all}

%\providecommand{\BIBYu}{Yu}

\end{document}